\documentclass[twocolumn, aps, floatfix]{revtex4}
\usepackage[T1]{fontenc}
\usepackage{helvet}
\usepackage{amssymb}
\usepackage{graphicx}

\begin{document}

\title{From a nonlinear string to a weakly interacting Bose gas}

\author{Emilia Witkowska$\,^1$, Mariusz Gajda$\,^{1, 2}$ and Jan Mostowski$\,^1$}

\address{$^1$ Institute of Physics, Polish Academy of Sciences, al.Lotnik\'ow $32/46$, 02-668 Warsaw, Poland}
\address{$^2$ Faculty of Mathematics and Natural Sciences, Card. Stefan Wyszy\'nski University,
	ul. Dewajtis $5$, 01-815 Warsaw, Poland}

\begin{abstract}

We investigate a real scalar field whose dynamics is governed by a nonlinear
wave equation. 
We show that classical description
can be applied to a quantum system of many interacting bosons provided that 
some quantum ingredients are included.
An universal action has to be 
introduced in order to define particle number. The value of this action should be equal
to the Planck constant.
This constrain can be imposed by removing high frequency modes from the dynamics by introducing a cut-off.
We show that the position of the cut-off has to be carefully
adjusted.
Finally, we show the proper choice of the cut-off 
ensures that all low frequency eigenenmodes which are
taken into account are macroscopically occupied.

\end{abstract}

\maketitle

\section{Introduction}

Description of interacting many body quantum system 
it is a very difficult task. Except of a few rather academic
problems exact solutions are not accessible and some approximated
methods are necessary. 
Recent attempts of description of a Bose-Einstein condensate at
finite temperature \cite{classical field_1}-\cite{Crit. temp.} showed that it is possible to
significantly simplify theoretical methods of solving
quantum dynamics.
This is due to the assumption of macroscopic occupation of single particle modes. 
It is worth noticing that Bose-Einstein condensate is not the only system
with macroscopic occupation of quantum states.

Historically the
first and the best known system with such properties is electromagnetic field.
As long as intensity is large classical approach based on Maxwell's equations
is valid.
The classical point of view was the only one used until the beginning of the twentieth
century. 
However this approach turned out to be inadequate to describe experiments
with small intensities and therefore the concept of photon had to be introduced.
On the other hand, physical phenomena involving
macroscopically large field amplitudes (or equivalently large number of photons) are successfully described
by the classical electric and magnetic fields
\cite{Maxwell field}.
Although quantization of electromagnetic field is a well established procedure,
the inverse procedure, i.e. substitution of a quantum field by a 
classical one is often heuristic and based on physical intuition rather
than formal arguments. At this point we should mention 
the Glauber theory of coherent states \cite{Glauber} of electromagnetic field.
If the field is in a coherent state then
electric and magnetic filed operators can be substituted by their 
non-vanishing mean values in this state. 
These mean values are interpreted as classical fields.

Although coherent states provide a link between 
classical and quantum theories the situation is not that simple
in case of particles with non-zero mass. Superselection rules
do not allow for superposition of states with different number of particles.
Therefore, one cannot introduce coherent states for such fields.
Such a situation takes place in the case of atomic Bose condensates,
where the number of particles is fixed.

In Bose condensates
at low temperatures only few lowest energy levels are macroscopically occupied.
One might expect that discrete structure of the matter field is not
essential and description of the system by a `classical wave' should be 
valid.
This expectation has no rigorous justification.
However an ingenious idea of Bogoliubov consisting in substitution of the annihilation
operator of a particle in the condensate mode by a c-number amplitude \cite{Bogoliubov}
is extremely successful and widely used. 
This approach leads to a
mean field description of the system in terms of a classical fields satisfying the Gross-Pitaevskii equation.

Thus the standard
theory of Bose condensate at zero temperature
is based on the Bogoliubov method.
Recently this idea was extended to finite temperatures
and is called the classical fields method.
It is successful in describing equilibrium properties of Bose 
condensate at finite temperatures, excitations spectrum, dissipative dynamics 
of vortices and many other
finite temperature phenomena \cite{classical field_2}.

The classical fields method is based on heuristic substitution 
of Bose operators by c-number amplitudes.
This substitution can be easily justified at very low temperatures since
practically all particles occupy the condensate mode.
On the other hand it seems 
questionable at temperatures close to the critical temperature.
Nevertheless the classical fields method works quite well up to the condensation point.
It is known that the Bose-Einstein condensation is a quantum phenomenon 
and the critical temperature depends on the Planck constant.

The Planck constant is usually introduced into theory through commutation relations for the field operators.
Those however are violated in the classical fields approach.
Nevertheless this approach preserves some bosonic features of the system.
Therefore it is justified to ask which features of the quantum system
are taken into account in the classical field methods and which assumptions leads to 
correct prediction of the condensation temperature.
In particular one may ask how does the Planck constant enter into the classical field method.

In order to answer these questions, at least partially,
we choose the following approach.
We start with a classical system of interacting harmonic 
oscillators and study its dynamics.
At a later stage we include some quantum ingredients and
pinpoint the moment where the Planck constant appears.
This approach justifies the classical fields method and shows its limitations.

The paper is organized as follows. In Sec. II we introduce the
model. We discuss the numerical techniques that we use
and main features of numerical solutions.
A particular form of interactions between oscillators leads to nonlinear equations 
similar to those studied by Fermi, Pasta and Ulam \cite{Fermi}.
Unlike 
the Fermi-Pasta-Ulam results the dynamics given by our model leads to thermalization of the system.
In Sec. III we analyze the state of thermal
equilibrium reached by the system, we study 
energy equipartition, and define the temperature of the system. 
In Sec. IV we analyze the system in terms of quasiparticles and
occupation of single particle state.
We show that elementary excitations (phonons) are distributed 
according to the low frequency part of Bose statistics.
High frequency part is introduced by hand with the help of properly chosen cut-off.
In this way we mimic quantum statistics in the whole range of frequencies.
In sec. V we conclude by summarizing our results and showing their implications for 
the classical fields method.

\section{Dynamics of a nonlinear string}

In this section we are going to introduce the model and its
basic equations. 
We consider a one dimensional elastic string of length $L$
and linear mass density $\rho$. In order to find dynamical equations of
motion we divide the string into $N-1$ elements of length $l_0=L/N$
and mass $m=l_0 \rho$. Each element is replaced by a point-like 
particle interacting with the nearest
neighbor via harmonic forces. The restoring force $F$ acting
on each particle is proportional to displacement $\Delta l_{0}$ from its equilibrium position, 
$F = - Y(\Delta l_0/l_0)$ where $Y$ is the
Young modulus. 
Thus the string can be viewed as $N$ particles moving on a line,
each of them connected to two neighbors by a spring with
equilibrium length $l_0$ and elastic constant $K=Y/l_0$.
We denote equilibrium positions of each particle (oscillator)
by $x_j=jl_0$ ($j=1,\dots,N$) and their displacements from
equilibrium (along the axis of the string) by $\phi_j$.
The Newton equations of motion for the displacements are:
\begin{equation}
m \ddot{\phi}_{j} = - K (2 \phi_{j} - \phi_{j+1} - \phi_{j-1}).
\label{L N E}
\end{equation}
For the future convenience we assume periodic boundary conditions,
i.e. $\phi_j=\phi_{j+N}$.
Eq. (\ref{L N E}) is used in different areas of physics, e.g. in description 
of vibration
of one-dimensional crystal lattice.
Analytic solutions of (\ref{L N E}) are available
in terms of plane waves, \cite{linear field}.

Let us remark, that in the limit of continuous medium, $N \rightarrow \infty$ (i.e. $l_0 \rightarrow 0$)
 Eq.(\ref{L N E}) takes the form of wave equation:
\begin{equation}
\frac{\partial^{2} \phi(x, t)}{\partial t^{2}} - c^2 \frac{\partial^{2}
\phi(x, t)}{\partial x^{2}} = 0 ,
\end{equation}
where $c=\sqrt{Y/\rho}$ is the velocity of sound.
In the  language of classical field theories equation (1)
describes free scalar field of zero charge.
In the following we will use the discrete version of the model.

We will now 
take into account a nonlinearity. 
For simplicity we assume that the nonlinear interaction is of short range (local)
and the dynamical equation is:
\begin{equation}
m \ddot{\phi}_{j} = - K(2 y_{j} - \phi_{j+1} - \phi_{j-1}) - \Lambda \phi_{j}^{3},
\label{nonlinear}
\end{equation}
where $\Lambda$ is a real parameter. 
This form of interaction
is widely used in various areas of physics,
in particular in the so called $\phi^4$ field
theory, \cite{phi4 theories}.

Equation (\ref{nonlinear}) is very similar to the one which appears 
in the famous Fermi-Pasta-Ulam (FPU) problem \cite{Fermi}. 
There are, however, two differences. First,
in the FPU  case displacements of the first and the last oscillators are
set to zero (as opposed to periodic boundary conditions assumed here).
Secondly, the nonlinear term in the FPU equation is of a different form.
The authors considered non-local nonlinear forces, for example 
of the form $(\phi_{j}-\phi_{j-1})^r$, where $r=2$
or $r=3$. The results of FPU calculations show that the system
shows
`very little, if any,
tendency toward equipartition of energy among degrees of freedom',
\cite{Fermi}.
On the contrary, as we are going to show in the following, the system
described by Eq.(\ref{nonlinear}) reaches a state of
thermal equilibrium characterized by equipartition of energy.

Let us introduce natural units: 1) unit of length $L$, 2) unit of
time $t_{0}=l_0/c$ and 3) unit of energy $\epsilon = K L^2$.
The set of coupled nonlinear equations takes the form:
\begin{equation}
\ddot{\phi}_{j} = - (2 \phi_{j} - \phi_{j+1} - \phi_{j-1}) - \lambda \phi_{j}^{3},
\label{Nn N E}
\end{equation}
where
\begin{equation}
\lambda=\frac{\Lambda}{K} L^2
\end{equation}
is the nonlinear coupling constant. 
We assume that $\lambda$, does not depend on $l_0$.
Therefore coefficients in
Eq.(\ref{Nn N E}) do not depend on the number of oscillators $N$.

Eq. (\ref{Nn N E}) leads to energy conservation:
\begin{equation}
E = \frac{1}{2} \sum_{j} [ \dot{\phi}_{j}^{2} + (\phi_{j}-\phi_{j-1})^{2} +
\lambda \frac{1}{2} \phi_{j}^{4} ]=const.
\label{H}
\end{equation}

We solve the set of equations (\ref{Nn N E}) numerically
for various initial energies and various number of oscillators $N$.
All initial
displacements $\phi_j$ and velocities $\dot{\phi}_j$ are generated from uniform probability
distribution on the interval $[-\varphi,\varphi]$,
where $\varphi$ is a parameter. Its value has to be adjusted according
to the value of initial energy. In general, the larger
initial energy the larger the value of $\varphi$.
In our numerical simulations we adjusted the time step, and therefore
accuracy of numerical procedure, to ensure energy conservation.

Due to assumed boundary
conditions it is convenient to analyze the results of our simulations
in the basis of plane waves
\begin{equation}
\phi_{j}(t)=\frac{1}{\sqrt{N}} \sum_{k=k_{min}}^{k_{max}} b_{k}(t) e^{-i k j},
\label{Ex}
\end{equation}
where the dimensionless wave vector $k$ takes the values
$k = n \pi/2$ and $n = - N, \dots, N-1$. 
Note that modes with $|k|> k_{max}$ are not present in (\ref{Ex}). The cut-off
of large $|k|$ is introduced in our model by discretization of space. We will
show that the cut-off position (or equivalently the number of spatial grid points $N$)
plays a very important role in our description.
Complex amplitudes of
different plane waves $b_k$ are subject to a constrain:
\begin{equation}
b_k(t)=b_{-k}^*(t),
\end{equation}
because the displacement field $\phi_j(t)$ is a real function.
This condition is automatically satisfied in numerical implementation.

\section{Thermal equilibrium}

The most important observation which follows from numerical solutions 
is that the system reaches a state of equilibrium
after some transient time.
 We checked this
observation by choosing many different initial conditions
corresponding to the same initial energy. The equilibrium state is
characterized by randomly-looking oscillations of each plane wave
amplitude $b_k(t)$.
Its modulus $|b_k(t)|^2$ oscillates in time from zero to some
maximal value. 
The mode with $k=0$ has the largest amplitude. In general, the larger the wave vector $|k|$,
the smaller the amplitude of oscillations of the corresponding mode.

\begin{figure}
\centering \resizebox{3.0 in}{6.0 in}
{\includegraphics{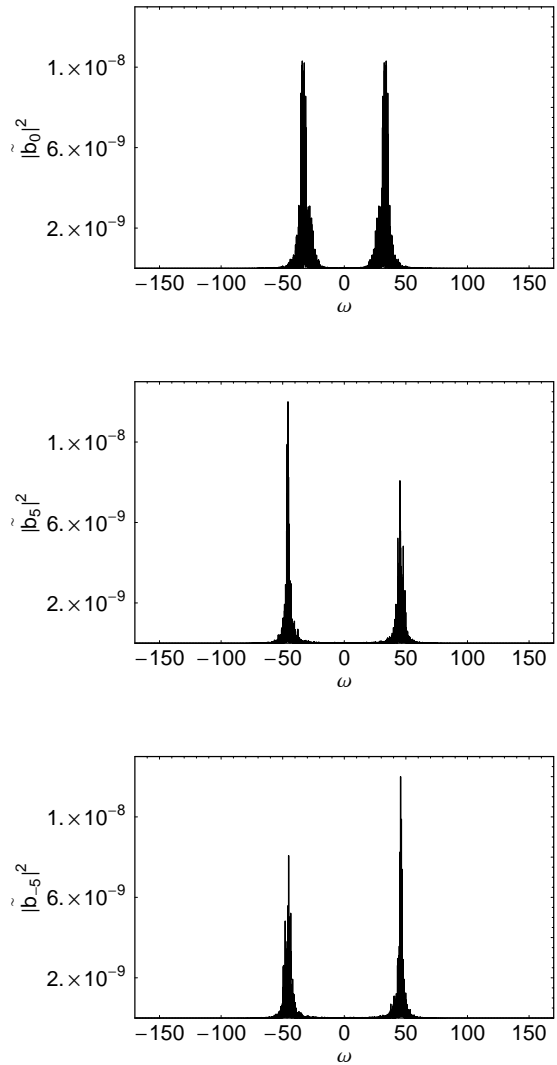}}
\caption{
Spectrum of amplitudes for energy $E = 0.1$, $\lambda = 1$, number of
grid points $N=64$ and a) $k = 0$, b) $k = - 5$, c) $k = 5$.
Note existence of two peaks centered at $\omega_{k}$
and $-\omega_{k}$.}
\label{widmo}
\end{figure}

We will now analyze the stationary state. 
It depends on the energy  $E$ and the number $N$.
Let us analyze the frequency spectrum of each plane wave amplitude 
$\tilde{b}_{k}(\omega)=\int dt e^{i \omega t} b_{k}(t)$.
It turns out that the randomly looking oscillations of $b_{k}(t)$ have a very regular spectrum.
Typical frequency spectra of plane wave amplitudes with $k = 0, -5, 5$ are plotted in Fig. \ref{widmo}.

The spectrum of each mode is composed of two peaks centered
around $\omega_k$ and $-\omega_k$. The peaks have finite width which
is relatively small as compared to the central frequency. Therefore,
in the first approximation, it is justified to neglect the width of peaks 
in the frequency spectrum
and assume that each plane wave amplitude oscillates
in time with two frequencies.
The frequency $\omega_k$ is defined as the weighted mean value of frequencies 
corresponding to the positive 
frequency peak:
\begin{equation}
\omega_{k} = \frac{\sum_{\omega>0} \omega |\tilde{b}_{k}(\omega)|^2}{\sum_{\omega>0} |\tilde{b}_{k}(\omega)|^2}
\end{equation}
while effective amplitude $\beta_k$ is:
\begin{equation}
\beta_{k} = \sqrt{\sum_{\omega>0} |\tilde{b}_{k}(\omega)|^2}
\end{equation} 
In this way we defined
an unique frequency $\omega_{k}$ and amplitude $\beta_{k}$ for every mode $k$.
Thus the amplitude $b_{k}$ has the following time dependence:
\begin{equation}
b_k(t)=\beta_k e^{-i \omega_k t}+\beta_{-k}^* e^{i \omega_k t}.
\end{equation}

\begin{figure}
\centering \resizebox{3.0 in}{1.9 in}
{\includegraphics{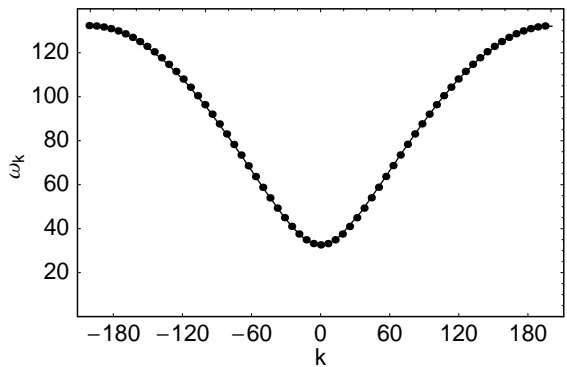}}
\caption{
Dispersion relation. Points refer to numerical
simulation with $E = 0.1$, $N=64$ and $\lambda = 1$. Curve refers to formula (\ref{F})
with $\alpha=4.5$.}
\label{dyspersja}
\end{figure}

We checked that the frequencies $\omega_{k}$
fulfill the following dispersion relation
\begin{equation}
\omega_{k} = \sqrt{4 \sin^{2}{\frac{k}{2}} + \omega^{2}_{0}},
\label{F}
\end{equation}
where $\omega_{0}$ is the frequency of the $k=0$ mode.
Numerical fit shows that:
\begin{equation}
\omega_{0}^{2} = 2 \lambda \alpha \sum_{k=k_{min}}^{k_{max}} |\beta_{k}|^{2},
\label{F0}
\end{equation}
and $\alpha$ is close to $4.5$.
In Fig. \ref{dyspersja} we show the dispersion relation
obtained from numerical simulations (points) and compare it to the 
analytic formula (\ref{F}).
Comparison shows remarkably good agreement.

In the continuous limit, where Eq. (\ref{Nn N E}) becomes a nonlinear 
wave equation the $4 \sin^{2}(k/2)$ term becomes simply
$k^2$. This difference is due to discretization of space
and the simplified form of the discrete version of the second spatial derivative. 

In the state of thermal equilibrium the dispalacment $\phi_{j}(t)$ 
can be approximated by
\begin{equation}
\phi_{j}(t)=\frac{1}{\sqrt{N}}\sum_{k=k_{min}}^{k_{max}} (\beta_k e^{-i (k j - \omega_k t)}
+ \beta_{-k}^* e^{i (k j - \omega_k t)}).
\end{equation}
Normal modes of the system are plane waves.
Every mode $k$ oscillates with two opposite frequencies
$\omega_k$ and $-\omega_k$. Amplitudes corresponding to these frequencies 
are related, $|\beta_{k}| = |\beta_{-k}|$. 
Using the language of the field theory
we can say that 
the positive frequency part corresponds to particle-like modes while the negative
frequency component corresponds to antiparticle-like excitations.

The total energy of the interacting system can be
 expressed in terms of amplitudes $\beta_{k}$ and
frequencies $\omega_{k}$ similarly as in the linear case:
\begin{equation}
E = 2 \sum_{k=k_{min}}^{k_{max}} \omega_{k}^{2} |\beta_{k}|^{2}.
\label{H A}
\end{equation}
Factor two which appears in the above formula comes from the fact
that each plane wave oscillates with two frequencies of opposite sign
and each of them gives the same contribution to the energy.
We check that disagreement between values of energy given by
Eq. (\ref{H}) and (\ref{H A}) is not larger than $5\%-10 \%$. Therefore, the total
energy is a sum of energies of independent modes.
The hamiltonian expressed in terms of plane waves is diagonal.

\begin{figure}
\centering \resizebox{3.0 in}{1.9 in}
{\includegraphics{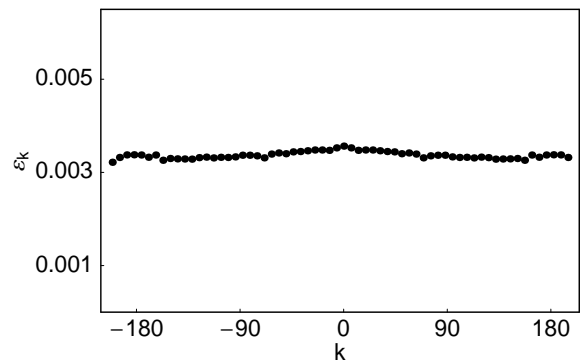}}
\caption{
Equipartition of energy. Energy per mode, Eq. (\ref{ekwipartycja}),
as a function of wave vector. Points refer to numerical simulation of Eq. (\ref{Nn N E})
with $E = 0.2$, $N=64$ and $\lambda = 1$.}
\label{ekwipartycja_fig}
\end{figure}

One of the most important results of
our numerical calculations is that total energy is evenly distributed
among all plane wave modes. The equilibrium state reached during
nonlinear evolution is characterized by equipartition of energy:
\begin{equation}
\varepsilon_{k} = 2 \omega_{k}^{2} |\beta_{k}|^{2} = const.
\label{ekwipartycja}
\end{equation}
where $\varepsilon_{k}$ is the energy per mode. 
Obviously $\varepsilon_{k}= E/N$.
Fig. (\ref{ekwipartycja_fig}) illustrates the equipartition of energy.
Let us remind that no equipartition of energy is observed in FPU problem.

Having the equipartition of energy we can define a temperature
of the system:
\begin{equation}
\tilde{T} = \varepsilon_{k},
\label{ekwipartycja 1}
\end{equation}
where temperature is expressed in unit of $\epsilon/k_{B}$ ($k_{B}$ is
the Boltzmann constant).

\section{Phonons}

In the previous section we have shown that the nonlinear hamiltonian (\ref{H}) 
can be expressed in the diagonal form (\ref{H A}) for each particular state of thermal
equilibrium. Note that it cannot be done `globally', i.e. independently
of the total energy of the system. This is  because the eigenfrequencies
depend on the energy.

The diagonal hamiltonian (\ref{H A}) is a sum of energies of $N$ independent 
harmonic oscillators
of frequencies $\omega_{k}$.
The interaction strength $\lambda$ enters 
this hamiltonian through the eigenfrequencies $\omega_{k}$ only.

This picture is purely classical.
In what follows we will reformulate our results in the language of quantum many body theory.
Our goal is to reach a better
understanding of foundations of the classical fields method.

The hamiltonian  (\ref{H A}) can be quantized with the help of canonical 
quantization procedure, i.e. by defining position and momentum in terms of amplitudes $\beta_{k}$,
and then imposing canonical commutation relation between them.
However, bosonic 
commutation relations are violated in the classical fields method -- classical amplitudes do commute.
Thus we will choose a different approach.

First we will define dimensionless amplitudes. 
These amplitudes are the only dynamical objects
appearing in the classical field method.
In order to define them we have to rewrite Eq.(\ref{H A}) 
using quantities with their proper dimensions:
\begin{equation}
\label{z jednostkami}
{\cal H}=m \sum_{k=k_{min}}^{k_{max}} \Omega_k^2 |A_{k}|^2, 
\end{equation}
where $\Omega_k=(1/t_0) \omega_k$ is the frequency and $A_k=\sqrt{2}L\beta_k$ is
a classical amplitude of each harmonic oscillator. 
To define dimensionless
amplitudes we need some {\it universal} quantity which has dimension of an action. 
Such a quantity does not exist in the classical theory. Therefore we 
have to introduce this constant by hand.
We will use symbol $\hbar$ for this elementary action.
At this moment its value can be arbitrary.
So far we have not mentioned
any physical condition which could, at least in principle, determine its value.
At a later stage this constant will be identified with the Planck constant.

Let us define units of amplitudes:
\begin{equation}
\label{action}
A_0(k)=\sqrt{\frac{\hbar}{m \Omega_k}}
\end{equation}
and dimensionless amplitudes for each oscillator:
\begin{equation}
B_k= \frac{L \sqrt{2}\beta_{k}}{A_{0}(k)} = \sqrt{\frac{2m\Omega_k}{\hbar}}L\beta_k,  
\label{amplitudes}
\end{equation}
and express the Hamiltonian Eq.(\ref{z jednostkami}) in terms of $B_k$:
\begin{equation}
\label{q hamiltonian}
{\cal H}= \sum_{k_{min}}^{k_{max}}\hbar \Omega_k |B_k|^2.
\end{equation}
This form of the Hamiltonian Eq.(\ref{q hamiltonian}) is a familiar one. 
The energy of each
plane wave mode is equal to the energy of elementary quantum of a given
frequency $\Omega_k$ times some positive real number:
\begin{equation}
\label{phonons}
N_k=|B_k|^2,
\end{equation}
which can be called a number of phonons. 
We want to stress once more that
number of phonons defined above is not an integer.
Therefore it can have 
a physical meaning of the number of quasiparticles only if its value is large as 
compared to one. Moreover, the value of $N_k$
depends still on the (arbitrary) value of the elementary action $\hbar$. 
Because there is no limitation on the maximal amplitude of 
a harmonic oscillator, the value of $N_k$ can be arbitrarily large. Our
classical field cannot therefore correspond to particles obeying
fermionic statistics.
However it can correspond to highly excited Bose field.

Equipartition of energy Eqs. (\ref{ekwipartycja}), (\ref{ekwipartycja 1}), discussed in the previous section,
can be expressed in terms of $N_{k}$:
\begin{equation}
\hbar \Omega_k N_{k} = k_{B} T,
\label{ekwi}
\end{equation}
where $k_{B}$ is the Boltzmann constant and $k_{B} T = \epsilon \tilde{T}$ where $\tilde{T}$
is known from our numerical simulations.
On the other hands 
if $N_k$ corresponds to the number of phonons (in the `classical
limit') than in the thermal equilibrium it should obey Bose statistics:
\begin{equation}
N_k=\frac{1}{e^{\hbar (\Omega_k-\mu)/k_{B} T}-1},
\end{equation}
where $\mu$ is a chemical potential.
In the limit of low frequencies $\Omega_k$, the equilibrium occupation of the mode $k$ 
can be approximated by:
\begin{equation}
N_{k} \hbar (\Omega_{k} - \mu) = k_{B} T.
\label{temp}
\end{equation}
Comparison of Eq. (\ref{temp}) and Eq. (\ref{ekwi}) shows that 
occupation of normal modes obtained in our calculations agrees with a low frequency 
limit of Bose statistics  provided that $\mu = 0$.

Equipartition of energy defines the energy per mode. 
All modes with wave vectors $|k|<k_{max}$ are occupied by phonons
while other modes are empty because of the momentum cut-off used in the implementation
of the model. This distribution of energy is
a $1D$ analogue of the Planck distribution of the blackbody radiation.
The Planck distribution says that energy density grows with frequency  up to some maximal value
and then falls exponentially to zero.
This initial growth is related to the
phase space volume which increases with $k$ as $\propto k^2$ and is
absent in the $1D$ model studied here. The exponential decay of the blackbody energy density
in replaced by a sharp cut-off at $k = k_{max}$ in our calculations.

We will use the similarities described above to determine the value of $\hbar$ which will allow
us to find the absolute value of the number of phonons $N_{k}$. 
Following the Planck idea we determine the value of 
$\hbar$ by equating the temperature of the system to the energy corresponding to the position of the maximum in the energy distribution. In our case this `maximum' corresponds to the cut-off:
\begin{equation}
\hbar \Omega_{k_{max}} = k_{B} T.
\label{hbar}
\end{equation}
Eq. (\ref{hbar}) plays a crucial role in establishing a link between classical fields
approach and the quantum theory of Bose system. It gives the value of the Planck constant up to a numerical
factor of the order of one. This fact establishes a limitation of accuracy of the method, in particular
the accuracy of the number of particles, or equivalently the value of the temperature of the system.
By comparing Eq.(\ref{hbar}) and Eq.(\ref{ekwi}) we see that number of particles occupying the
mode with the largest energy is equal to one:
\begin{equation}
N_{k_{max}} = 1.
\label{cut-off}
\end{equation}
This means that occupation of all modes with $k$ smaller than $k_{max}$ 
is macroscopic, $N_k>1$, which is in agreement with
a common understanding of the classical limit of the quantum field. Indeed in the classical
fields method the number of modes used in numerical implementation has to be carefully adjusted.
As suggested in \cite{Zawitkowski} it should be such that the occupation of the highest mode is equal to 1.

\begin{figure}
\centering \resizebox{3.1 in}{1.9 in}
{\includegraphics{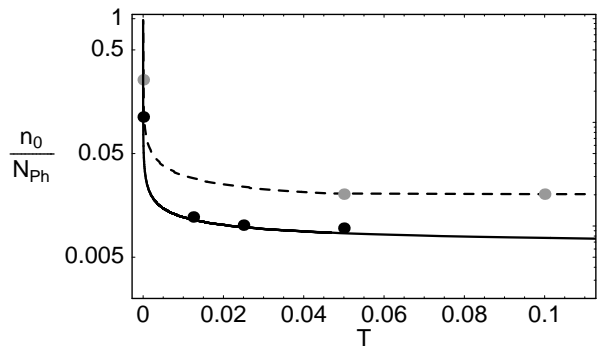}}
\caption{Relative number of phonons in $k=0$ mode as a function of temperature
for different total number of
quasiparticles: $N_{Ph} = 85$ (gray circles) and $N_{Ph} = 175$ (black circles). 
Points refer to numerical simulations of Eq.(4) with $\lambda = 1$.
Lines represent a fit with $\alpha = 4.5$.}
\label{1D}
\end{figure}
In what follows we give an illustrative example of application
of the defined above procedure of `quantization of a scalar field' studied in this paper.
We solve the dynamical equations for various total energy of the system and various
number of grid points $N$. In our calculations we used rather small values of $N$
ranging from $N=16$ up to $N=1024$. 
We check if the system reached the state of thermal equilibrium by
determining a frequency spectrum $\omega_k$ and amplitudes $\beta_k$ at various stages
of time evolution. When the stationary state is reached we control  equipartition
of energy and determine the temperture. Finally, occupation of different eigenmodes
is obtained from the following relation:
\begin{equation}
N_k=\frac{|\beta_k|^2}{|\beta_{k_{max}}|^2}.
\label{occup}
\end{equation}
Evidently $N_{k_{max}}=1$, what is equivalent to the condition (\ref{hbar}).
If total number of phonons $N_{Ph}=\sum N_k$ is different than assumed we have to
repeat the calculations changing the number of grid points. In general the smaller
the energy the smaller $N$ has to be used in order to keep the same number
of phonons. This way we are able
to get occupations of all normal modes as functions of temperature for a fixed
total number of phonons.

In Fig.(\ref{1D}) we present occupation of $k=0$ mode as a
function of temperature for nonlinear 
interaction strength $\lambda = 1$ and two different total number of phonons.
The solid line represents a fit to the numerical results. The results
show that relative occupation of the spatially uniform $k=0$ mode grows quite
rapidly at temperatures close to zero. There is no phase transition in 
the studied system because, first of all, our system is one-dimensional, and
also elementary excitations are massless. The main goal of these 
calculations was to illustrate how the classical field method works in practice.   

\section{Concluding remarks}

In this paper we started from the description of a purely classical system. Our numerical simulations
proved that nonlinear interactions drive the system towards an equilibrium. Energy in the equilibrium 
is evenly distributed among all modes of $|k|\le k_{max}$. Modes corresponding
to $|k|> k_{max}$ are not excited and do not contribute to the total energy. They are
absent in our approach because of
the cut-off which is an essential ingredient of the method. In order to `quantize'
the dynamics we introduced the Planck constant.
Then we could define dimensionless amplitudes
of each oscillator and the number of excitation quanta of each mode -- phonons. However, the number of phonons
depends on the value of $\hbar$ which, at this stage of the approach, can be arbitrary. In order to
assign a value to $\hbar$ we followed the Planck's idea. We  used the fact that Planck's
distribution of energy leads to its equipartition for small values of $k$ and exponential
drop for large $k$. Energy distribution reaches a maximum at frequency
$\omega_{k_{max}}$ close to this satisfying
the relation $\hbar \omega_{k_{max}} = k_B T$. In our approach the same condition was used to determine
the value of $\hbar$ and thus the number of phonons $N_{Ph}$.

We believe that our studies shed more light on the classical fields method used for description
of a weakly interacting Bose-Einstein condensate. The way of reasoning 
goes the opposite way in `derivation' of this method. 
One starts with quantum many body theory of a Bose gas with short range
interactions. Two particle interaction energy is $g/V$ where $g$ is proportional to the s-wave 
scattering length and $V$ is the volume. The number of macroscopically occupied modes
is being assumed {\it a priori} by choosing a value of the cut-off momentum $k_{max}$. Annihilation
operators of these selected modes are then substituted by classical amplitudes $\alpha_k$
and Heisenberg operator equations transform into equations for classical amplitudes $\alpha_{k}$:  
\begin{equation}
i\hbar \frac{{\rm d}}{{\rm dt}} \alpha_k = \frac{\hbar^2 k^2}{2m} \alpha_k +\frac{gN_p}{V}\sum_{k_{1},k_{2}}
\alpha^{*}_{k_1} \alpha_{k_2} \alpha_{k_1-k_2+k},
\end{equation}
where $N_p$ is the total number of particles. This quantity corresponds to the number $N_{Ph}$ 
of phonons i the discussed classical model. Note that $N_p$ does not enter dynamical equations 
alone -- it is multiplied by the interaction strength $g$.

Relative occupation of different $k$ modes is closely related to the classical amplitudes:
\begin{equation}
|\alpha_k|^2=\frac{N_k}{N_p}=n_k.
\end{equation}
Let us observe that the Planck constant appears in the equations right from the beginning. But 
its value can be {\it arbitrary} -- the classical amplitudes $\alpha_k$ are a direct analogue of
amplitudes $B_k$ of our model. 

Dynamics of classical fields leads to the equipartition of energy: 
\begin{equation}
\hbar \omega_k n_k=\frac{k_B T}{N_p}=const.
\label{xxx}
\end{equation}
Only the value of $\hbar \omega_k n_k$ is known from numerical calculations, therefore
(\ref{xxx}) allows for determination of the ratio of $k_BT/N_p$ but not the values of $T$, $N_{p}$ separately.
Note that both $N_{p}$ and $g$ are not uniquely determined 
because only product $gN_p$ is an initial control parameter.
According to the present studies
this problem can be resolved if the value of $\hbar$ is determined by the requirement
that the energy distribution of classical fields mimics the quantum distribution -- the position of maximum
ought to be approximately given by 
\begin{equation}
\label{planck}
\hbar \omega_{k_{max}} = k_B T. 
\end{equation}
This equation together with the equipartition relation allows for determination of the particle number:  
\begin{equation}
N_p=\frac{1}{n_{k_{max}}}.
\label{cutoff}
\end{equation}
In the classical fields method the above condition is justified on a basis of heuristic arguments by
saying that
all classical modes have to be macroscopically occupied.
The approach we used gives a new interpretation of this reasoning. 

Moreover, our studies show limits of the `predictive power' of the classical fields method.
Because the Planck constant is determined with limited accuracy all predictions
about the total number of particles or temperature of the system are accurate up to a numerical
factor of the order of one. 
It seems therefore that attempts of more accurate determination of the temperature
\cite{Davis} cannot be free of the ambiguity related to the position of the cut-off.
In addition the classical fields method cannot be used for
description of very subtle effects such as a shift of the critical temperature
of Bose-Einstein condensation due to interactions. For a typical system this shift is very small.
Some efforts \cite{Crit. temp.} of obtaining the shift of critical temperature using classical fields method
must be  inevitably biased by the approximate character of Eq.(\ref{planck}).  

\acknowledgments

This work was supported by the Polish Ministry of Scientific Research and 
Information Technology under Grant No. PBZ-MIN-008/P03/2003.


\begin{thebibliography}{99}

\bibitem{classical field_1} 
K. G\'oral, M. Gajda, K. Rz\k{a}\.zewski, Opt. Express \textbf{8}, 92 (2001);
M. J. Davis, S. A. Morgan and K. Burnett
Phys. Rev. Lett. \textbf{87}, 160402 (2001);
A. Sinatra, C. Lobo, Y. Castin, J. Phys. B: At. Mol. Opt. Phys. \textbf{35}, 3599-3631 (2002);
A. Sinatra, C. Lobo, Y. Castin,
Phys. Rev. Lett. \textbf{87}, 210404 (2001).

\bibitem{classical field_2}
H. Shmidt, K. G\'oral, F. Floegel, M. Gajda, K. Rz\k{a}\.zewski
J. Opt. B \textbf{5}, S96 (2003);
M. Brewczyk, P. Borowski, M. Gajda, and K. Rz\k{a}\.zewski, J. Phys. B. \textbf{37}, 2725 (2004);
C. Lobo, A. Sinatra, Y. Castin, Phys. Rev. Lett. \textbf{92}, 020403 (2004),
D. Kadio, M. Gajda and K. Rz\k{a}\.zewski,
Phys. Rev. A, \textbf{72}, 013607 (2005).

\bibitem{Crit. temp.} M. J. Davis and P. B. Blakie, e-print: cond-mat/0508667.

\bibitem{Maxwell field} J. D. Jackson
\emph{Classical Electrodynamics}, New York: John Wiley and Sons, Inc. (1962);
I. Bia\l{}ynicki-Birula, Z. Bia\l{}ynicka-Birula
\emph{Quantum Electrodynamics}, Oxford - Warszawa: Pergamon Press - PWN (1975).

\bibitem{Glauber} R. J. Glauber `Quantum Optics and phonon statistics'
in \emph{Quantum Optics and electronic}, C. DeWitt, A. Blandin
and C. Cohen-Tannoudji, eds., Gordon and Breach, New York, 1965;
Phys. Rev. \textbf{130}, 2529 (1963);
Phys. Rev. \textbf{131}, 2766 (1963).

\bibitem{Bogoliubov} N. N. Bogoliubov, J. Phys. \textbf{11}, 23 (1947);
F. Dalfovo, S. Giorgini, L. P. Pitaevskii, S. Stringari,
Rev. Mod. Phys. \textbf{71}, 463 (1999).

\bibitem{Fermi} S. M. Ulam, E. Fermi, J. Pasta \emph{Studies of nonlinear problem},
Analogies between Analogies, University of California Press, 1990;
The summary review one can find in
G. P. Berman and F. M. Izrailev, e-print: cond-mat/0411062.

\bibitem{linear field}
The linear case is fully solvable, see:
C. Cohen-Tannoudji, B. Diu , F. Lalo\"e, \emph{Quantum Mechanics}, New York: J. Wiley, (1977);
N. N. Bogoliubov, D. W. Shirkov
\emph{Introduction to the Theory of Quantized Fields}, John Wiley and Sons Inc. (1980).

\bibitem{phi4 theories} L. Caiani, L. Casetti, C. Clementi, G. Pettini
and R. Gatto, Phys. Rev. E, \textbf{57}, 3886 (1998).

\bibitem{Zawitkowski} \L{}. Zawitkowski, M. Brewczyk, M. Gajda, K. Rz\k{a}\.zewski, 
Phys. Rev. A, \textbf{70}, 033614 (2004).

\bibitem{Davis} M. J. Davis and S. A. Morgan, Phys. Rev. A  \textbf{68} 053615 (2003);
M. J. Davis and P. B. Blakie, J. Phys. A: Math. Gen. \textbf{38} 10259 (2005).

\end{thebibliography}
\end{document}